\documentclass[prl,aps,twocolumn,showpacs,floatfix]{revtex4}
 \newcommand{\mytitle}[1]{
 \twocolumn[\hsize\textwidth\columnwidth\hsize
 \csname@twocolumnfalse\endcsname #1 \vspace{1mm}]}
 \newcommand{\beq}{\begin{equation}}
 \newcommand{\eeq}{\end{equation}}
 \newcommand{\bea}{\begin{eqnarray}}
 \newcommand{\eea}{\end{eqnarray}}
 \newcommand{\pdag}{{\phantom{\dagger}}}
\usepackage{graphics}
\usepackage{graphicx}
\usepackage{bm}

\begin{document}
\draft

\title{Kondo effect in coupled quantum dots 
with RKKY interaction:\\
Finite temperature and magnetic field effects}
\author{Chung-Hou Chung$^{1}$ and Walter Hofstetter$^{2}$}
\affiliation{
\mbox{$^{1}$Institut f\"ur Theorie der Kondensierten
Materie, Universit\"at Karlsruhe, 76128 Karlsruhe, Germany}\\
$^{2}$Institut f\"ur Theoretische Physik, Johann Wolfgang Goethe-Universit\"at, 
60438 Frankfurt/Main, Germany
}
\date{\today}

\begin{abstract}
We study transport through two quantum dots coupled by an RKKY interaction 
as a function of temperature and magnetic field.
By applying the Numerical Renormalization Group (NRG) method  we obtain 
the transmission and the linear conductance. 
At zero temperature and magnetic field, we observe a quantum phase 
transition between the Kondo screened state and a local 
spin singlet as the RKKY interaction is tuned.
Above the critical RKKY coupling  the Kondo peak is split. 
However, we find that both finite temperature and magnetic field 
restore the Kondo resonance. 
Our results agree well with recent transport experiments on gold grain quantum dots 
in the presence of magnetic impurities \cite{Heersche}.  
\end{abstract}

\maketitle

\emph{Introduction.} 
In recent years the Kondo effect \cite{hewson} in 
semiconductor quantum dots has gained 
significant interest both theoretically and 
experimentally \cite{Kondo-popular,kondo,kondo-theo}.  
Electronic transport in quantum dots is strongly influenced by the Coulomb 
blockade \cite{Coulomb-blockade} 
due to their small size. 
In these systems, a single unpaired spin can interact with conduction electrons, 
leading to screening of the spin and an enhanced 
conductance at low bias and low temperatures. 
More recently, due to rapid 
progress in spintronics and quantum information, it becomes desirable 
to gain more tunable spin control in double quantum dots 
where an effective spin-spin interaction known as the 
Ruderman-Kittel-Kasuya-Yoshida (RKKY)\cite{RKKY} coupling is generated between 
the two dots via conduction electrons in the leads. 
The RKKY coupling competes with the Kondo effect in these systems, 
leading to a quantum phase transition between the 
Kondo screened phase at weak RKKY coupling and a 
local spin-singlet state at strong coupling. 
Part of this rich physics has been studied previously 
in the two-impurity Kondo \cite{2impkondo} and Anderson \cite{2impAnderson} 
problems.
  
Recently, indications for a competition between Kondo screening and the RKKY interaction 
have been observed experimentally \cite{Craig}. 
This experiment stimulated theoretical and experimental 
efforts on the above-mentioned quantum phase 
transition in coupled dots. 
Very recently, new measurements have been reported 
to observe the restoring of the Kondo resonance in a gold quantum dot with 
magnetic impurities at finite temperatures and magnetic fields where 
an effective RKKY coupling is generated between the dot and the 
impurities \cite{Heersche}.  
Though similar systems have been studied theoretically via several  
approaches \cite{pre_2impRKKY1,pre_2impRKKY2,pre_2impRKKY3},
little is known about the transport properties 
at finite temperatures and magnetic fields. 
Motivated by these recent experiments, here we systematically study  
transport properties of double quantum dots coupled by an RKKY 
interaction at finite temperatures and magnetic fields 
via the Numerical Renormalization Group (NRG) method \cite{nrg}, 
a non-perturbative approach 
to quantum impurity systems. In contrast to 
other techniques, this method does not rely on any assumptions 
regarding the ground state or the leading divergent couplings, 
which is crucial in our analysis.

\begin{figure}[h]
\begin{center}
\includegraphics[width=0.7\linewidth]{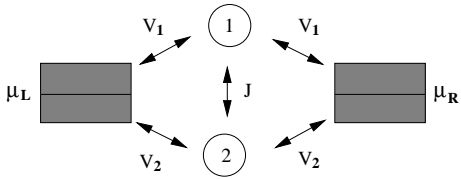}
\end{center}
\vspace{-0.5cm}
\caption{\label{set}Two quantum dots (denoted by the two circles) with 
two--channel leads (chemical potentials $\mu_L$ and $\mu_R$). 
$V_1$ and $V_2$ denote the tunneling matrix element between the dots 
and the leads. There exists an RKKY coupling $J$ between the two dots.}
\end{figure}

\emph{The Model.} 
We 
consider a two-impurity Anderson model as shown in Fig.~\ref{set}, 
describing two quantum dots which are seperately coupled to two-channel leads 
and subject to an antiferromagnetic RKKY spin-spin interaction and a local 
magnetic field. This setup is general enough to 
describe both experiments mentioned above.
The Hamiltonian of the system is given by:
\begin{eqnarray}
H &=& H_D + H_{\it l} + H_t +H_{J} +H_B, \nonumber \\
H_D &=& \sum_{i s} (\epsilon_{d i} + \frac{U}{2}) 
d^{\dagger}_{i\sigma}d_{i\sigma} + 
\frac{U}{2} \sum_i (N_i-{\mathcal N})^2, \nonumber \\
H_{\it l} &=& \sum_{\alpha i k \sigma}\epsilon_{\alpha i k \sigma} 
c^{\dagger}_{\alpha i k \sigma} c_{\alpha i k \sigma},\nonumber \\ 
H_t &=& \sum_{\alpha i k \sigma} V_{i \alpha} 
c^{\dagger}_{\alpha i k \sigma} d_{i \sigma} + h.c., \nonumber \\
H_{J} &=& J \, {\bf S}_1\,  {\bf S}_2,\,\,\,\,\,\, 
H_B= - B ({\bf S}_{1 z} + {\bf S}_{2 z})
\end{eqnarray}

Here, $\alpha=L/R$ denotes the left/right lead, 
$i=1,2$ denotes the two dots, 
$N_i=\sum_{\sigma} d^\dagger_{i\sigma} d^\pdag_{i\sigma}$ 
is the number of electrons occupying dot $i$ and 
${\bf S}_i = (1/2)\sum_{\sigma \sigma'} d^\dagger_{i\sigma}  
\bm{\sigma}_{\sigma \sigma'} d^\pdag_{i\sigma'}$
are the spins of the two levels. Each dot is subject to a   
 charging energy $U$ and $J$ is the  
RKKY exchange coupling between the two dots which is assumed to 
be antiferromagnetic ($J>0$). 
The magnetic field $B$ induces a Zeeman splitting, where we have set $g=1$. 
Here, we consider the model with particle-hole symmetry, i.e. 
$\epsilon_{d i} = -\frac{U}{2}$, and 
single occupation for each dot, i.e. ${\mathcal N} =1$. 
The energies $\epsilon_{di}$ of the two dots and their  
precise position in the Coulomb blockade valley can be tuned experimentally 
by an external magnetic field and the gate voltage.
Here, we neglect the energy dependence of the tunneling matrix elements 
$V_{i \alpha}$. Without loss of generality, we assume they have 
left-right symmetry, {\it i.e.} $V_{1L} = V_{1R}= V_1$, 
$V_{2L} = V_{2R}= V_2$, but $V_1\neq V_2$ in general. 
The intrinsic line width of the dot levels due to tunnel coupling to the leads 
is $\Gamma = \Gamma_L + \Gamma_R$ with $\Gamma_{L/R} = 2\pi |V|^2 N_{L/R}$ 
where $N_{L/R}$ is the density of states in the leads. 

For $J=0$, and $V_{i r}=0$, three triplet configurations 
$|1,1\rangle = d^\dagger_{1\uparrow} d^\dagger_{2\uparrow} |0\rangle$, 
$|1,0\rangle = (1/\sqrt{2}) (d^\dagger_{1\uparrow} d^\dagger_{2\downarrow} 
+ d^\dagger_{1\downarrow} d^\dagger_{2\uparrow} |0\rangle$, 
$|1,-1\rangle = d^\dagger_{1\downarrow} d^\dagger_{2\downarrow}|0\rangle$ 
and the singlet 
$|0,0\rangle = d^\dagger_{1\uparrow} d^\dagger_{2\downarrow} |0\rangle$ 
are degenerate. 
Finite tunneling $V_{i r}$ into the leads to independent spin-1/2 Kondo 
screening in each 
of the two dots. 
The corresponding Kondo temperatures $T_{K}^{1}$ and $T_{K}^{2}$ 
are generally different, and given by 
$T_{K}^{1(2)} \propto D \exp(-1/2\rho_c J_{1(2)})$ where $D$ is the bandwidth, 
$\rho_c$ the density of states of the leads, and $J_{1(2)} = 4 V_{1(2)}^2/U$ 
the effective Kondo coupling \cite{hewson}.

The above degeneracy is lifted at finite RKKY coupling $J$: three triplet 
states are shifted to energy $E_t=J/4$ and the singlet state to energy 
$E_s = -3J/4$. There exists 
competition between Kondo screening and a local  
spin-singlet ground state: the former is expected to be the ground state 
for large $J$, the latter for small $J$. 
A quantum phase transition at zero temperature 
between these two phases occurs as the RKKY coupling is 
tuned \cite{2impkondo,2impAnderson}. 
Note that a related singlet-triplet transition in two-level quantum dots 
was studied both for two-mode \cite{sakai2001} 
and for single-mode leads \cite{2level_1,2level_1_glazman,2level_2}. 
In the latter case, a Kosterlitz-Thouless transition 
from a local singlet state to a single-channel $S=1$ underscreened Kondo model 
was observed \cite{2level_1}. 

An interesting aspect of the Kondo-to-singlet 
transition in our setup is the tunability between 
these two phases since one can get good control over the various 
parameters in experiments. In particular, as observed in the experiment  
\cite{Heersche}, the Kondo resonance is restored 
at finite temperatures and magnetic fields 
close to the singlet-triplet transition. The goal of our work is to  
describe this behavior theoretically. 

\emph{Transport properties.}
We are interested in calculating electronic transport through 
the dot close to the transition. 
To this end, we use the generalized Landauer formula \cite{landauer} 
\beq \label{current}
I = \frac{2 e}{h} \int d\omega\, 
\left(f(\omega - \mu_L) - f(\omega - \mu_R)\right)\, T(\omega)
\eeq
with the Fermi function $f(\omega)$ and the transmission 
coefficient \cite{2level_1}  
\beq \label{transmission}
T(\omega) = - \sum_{i,\sigma} 
\frac{\Gamma^L \Gamma^R}{\Gamma^L + \Gamma^R}\;
{\rm Im} G_{ii \sigma}(\omega).
\eeq
Here we have introduced the retarded dot Green's functions 
$G_{i i' \sigma}(t) = -i \theta(t) \langle \{d^\pdag_{i\sigma}(t), 
d^\dagger_{i'\sigma}\} \rangle$. In the following we focus on the low bias 
regime, where $T(\omega)$ can be evaluated in equilibrium, using the NRG. 
For a detailed description of this technique see Ref.~\cite{nrg}.
Using the current formula (\ref{current}), we determine the behavior 
of the linear conductance $G(T)=\frac{dI}{dV}|_{V=0}$ 
at finite temperatures. Note that the equilibrium transmission 
$T(\omega)$ also yields a good approximation to the   
differential conductance $dI/dV$ at finite bias 
measured in experiments. 

\begin{figure}[h]
\begin{center}
\includegraphics[width=1.0\linewidth]{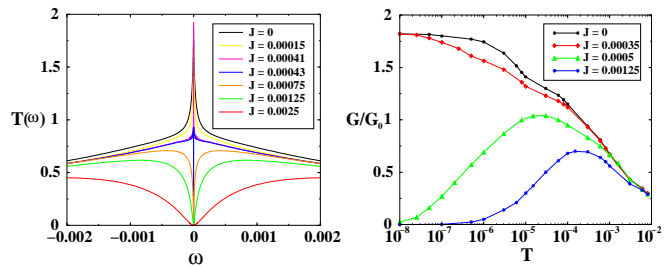}
\end{center}
\vspace{-0.5cm}
\caption{\label{TwJ}(a) Transmission coefficient at zero temperature 
for different RKKY interactions. The unit of energy is the half bandwidth $D=1$ of 
the conduction electrons. Here, $U=1$, $\epsilon_{d 1}=\epsilon_{d 2}=-0.5$
, $\Gamma_{1L}=\Gamma_{1 R}=0.05$, $\Gamma_{2L}=\Gamma_{2R}=0.1$, $\Lambda=4$, 
$T_k^{1}\approx 0.002$, $T_k^2 \approx 0.00002$ (for $J=0$). The critical 
RKKY coupling is $J_c\approx 0.00042$. (b) Linear conductance G(T) 
for different RKKY couplings. Parameters are the same as in (a).  
}
\end{figure}

\emph{Quantum phase transition at zero temperature for zero field.} 
In Fig.~\ref{TwJ}(a) we plot the transmission coefficient $T(w)$ 
as a function of RKKY coupling $J$ for zero temperature and zero field.  
It shows a quantum phase transition 
between the Kondo screened phase ($J<J_c$) 
and the local spin-singlet phase (for $J> J_c$), where $T_{k}^1<J_c<T_{k}^2$.
The value of $J_c$ is consistent with previous results by 
Sakai {\it et al.} \cite{2impAnderson}. 
For $J<J_c$, a Kondo peak at $\omega=0$ is observed, 
where the transmission $T(\omega\to 0)$ tends to reach 
its unitary limit of $2$ 
(each dot acts as an unitary Kondo channel). 
Due to systematic numerical errors 
in the NRG calculation, this limit is underestimated by about $10\%$.
As $J$ is increased, the Kondo peak becomes narrower, indicating 
a vanishing low-energy scale as the system approaches the critical point. 
As $J\to J_c$, the transmission reaches the unstable fixed point value 
$T(\omega\to 0) = 1$.   
For $J>J_c$, the two dots form a local spin-singlet state where the 
Kondo peak is split and $T(\omega\to 0)=0$. 
The splitting of the Kondo peak increases as $J$ is increased. 

\emph{Transport at finite temperatures.} 
We now determine the behaviour 
of the linear conductance $G(T)/G_0$ 
at finite temperatures where $G_0=2e^2/h$ is the conductance unit. 
Results are shown in Fig.~\ref{TwJ}(b). 
The Kondo screened and local spin-singlet 
phases are characterized by 
stable fixed points with $G(T\to 0)\approx 2$ or $0$, respectively. 
In the Kondo regime, upon lowering 
the temperature, the conductance rises up to the unitary limit 
with two step-like structures indicating the two crossover 
Kondo temperatures $T_{k}^1$ and $T_{k}^2$. 
%
%

In the local spin-singlet regime we 
find a nonmonotonic behaviour of the conductance 
when $T$ is lowered: After an initial rise due to the Kondo effect, 
$G(T)$ decreases to zero as the temperature is lowered. The broad 
peak of $G(T)$ at finite temperatures is in fact a signature of 
the reappearance of the Kondo resonance which is also seen in the experiment \cite{Heersche}. 

\begin{figure}[h]
\begin{center}
\includegraphics[width=0.72\linewidth]{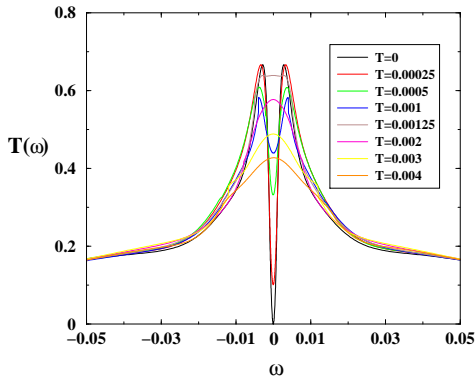}
\end{center}
\vspace{-0.5cm}
\caption{\label{TwT}Transmission coefficient $T(w)$ for a fixed 
RKKY coupling $J=0.002$ at finite temperatures. Here, 
$\Gamma_{1L}=\Gamma_{1 R}=0.04$, $\Gamma_{2L}=\Gamma_{2R}=0.08$. 
The remaining parameters are the same as in Fig. 2. 
}
\end{figure}
To gain more insight into the finite temperature behavior, 
we present the plot of transmission $T(\omega)$ at different temperatures 
(see Fig.~\ref{TwT}) when
RKKY is strong enough ($J > J_c$) so that the Kondo peak is split at zero 
temperature.
We find that as temperature increases, $T(\omega)$ 
inside the dip first increases due to thermal broadening of the 
split peaks around $\omega\approx \pm J$  
until the Kondo resonance reappears. Then the peak height decreases again as $T$
is further increased, similar to the Kondo effect at finite temperatures 
without RKKY interaction. The transmission $T(\omega=0)$ 
reaches its maximum value at a temperature $T_{\rm max} \propto J - J_c $. 
Note that the NRG data for $T(\omega)$ 
at $\omega\le T$ are estimated by extrapolating $T(\omega)$ 
from higher frequencies. 
The qualitative behavior of the transmission -- reappearance of the 
Kondo peak at finite $T$ --  agrees well with recent experiments \cite{Heersche}. 

\begin{figure}[h]
\begin{center}
\includegraphics[width=1.0\linewidth]{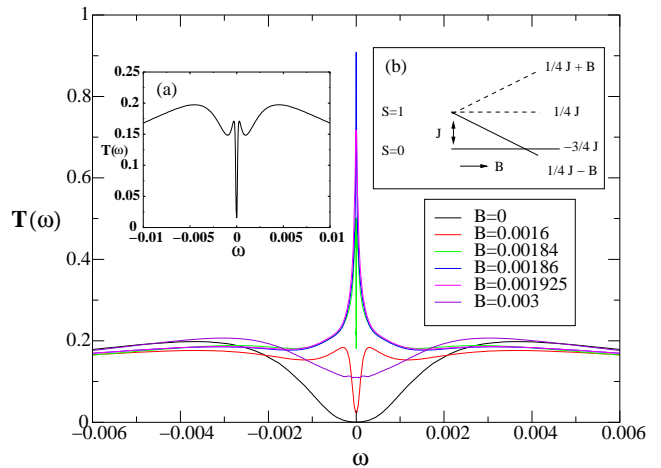}
\end{center}
\vspace{-0.5cm}
\caption{\label{TwB} Transmission coefficient $T(w)$ of the coupled 
dots at different magnetic fields. Here, 
$\Gamma_{1L}=\Gamma_{1 R}=\Gamma_{2L}=\Gamma_{2R}=0.06$, $J=0.002$. 
The remaining parameters are the same as in Fig.~\ref{TwJ}. Inset (a): 
the transmission $T(\omega)$ for $B=0.0016$, which clearly shows 
four peaks in pairs around $\pm(B\pm J)$. Inset (b): The energy levels 
of the triplet and singlet states in the presence of a finite RKKY 
coupling $J$ and a magnetic field $B$. The three triplet states are split  
into $E_{1,1}=1/4 J-B$, $E_{1,0}=1/4 J$, and $E_{1,-1}=1/4 J+B$. 
The singlet $|0,0\rangle$ and one of the triplet states $|1,1\rangle$ 
become degenerate at $B=J$.
}
\end{figure}

\emph{Effect of a magnetic field.} 
In the presence of a magnetic field a singlet-triplet crossover occurs. 
Here we discuss the effect of a Zeeman splitting at zero temperature 
in the presence of a large RKKY coupling $J$ such that 
the Kondo peak is split in the absence of magnetic fields 
(see Fig.~\ref{TwB}). 
For finite $B < J$, 
we find an increase of the transmission $T(\omega)$ 
at $\omega=0$. When $B$ is comparable to the value of RKKY interaction, $B\approx J$, 
we observe the reappearance of the Kondo peak where $T(\omega)$ 
reaches the unitary limit of $1$, corresponding to 
a single channel $S=1/2$ Kondo effect. When the magnetic 
field increases further, the Kondo peak splits again.  
%
%
 What happens is that due to the Zeeman splitting, one
component of the triplet ($|1,1\rangle$) is "pulled down" and 
eventually becomes degenerate with the singlet $|0,0\rangle$ 
(see inset (b) of Fig.~\ref{TwB}). 
At this point a Kondo effect -- analogous to $S=1/2$ Kondo --
arises between these two states, which has been discussed previously 
for a 2-level quantum dot 
within a perturbative scaling approach \cite{STGlazman}.
Our nonperturbative NRG results confirm this scenario: 
The transmission $T(\omega=0)$ indeed 
reaches the unitary limit of the $S=1/2$ Kondo effect. 
Note that due to tunneling into the leads, degeneracy of the singlet and triplet state 
occurs at a renormalized value $B \ne J$. 
Close to the degeneracy point, a four-peak structure emerges in the 
transmission (see inset (a) of Fig.~\ref{TwB}), 
corresponding to the splitting between the singlet 
and the lowest (second-lowest) triplet state.  
Due to NRG broadening effects, the third triplet state 
is not visible.

\begin{figure}[h]
\begin{center}
\includegraphics[width=1.0\linewidth]{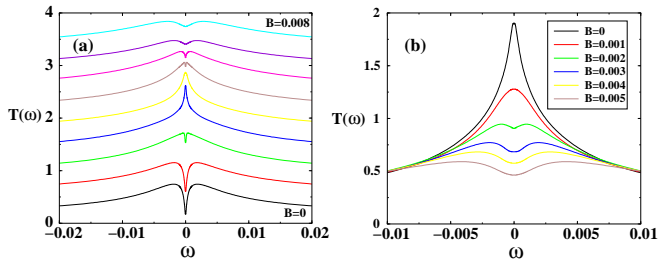}
\end{center}
\vspace{-0.5cm}
\caption{(a) Transmission coefficient $T(\omega)$ 
of the coupled quantum dots for different magnetic fields. 
Here, $U=1$, $\epsilon_{d 1}=\epsilon_{d 2}=-0.5$
, $\Gamma_{1L}=\Gamma_{1 R}= \Gamma_{2L}=\Gamma_{2R}=0.1$, $\Lambda=4$, 
$J=0.007$, $T_k \approx 0.0025$ (for $J=0$), $J_c\approx 0.005$, $T=0.00001$. 
The values of $B$ are in steps of $0.001$, and the traces of  
$T(\omega)$ are shifted in steps of $400 B$. (b) Transmission vs. frequency  
for ferromagnetic RKKY coupling $J=-0.005$
and different magnetic fields. The remaining parameters 
are the same as in (a).
\label{TwB2}}
\end{figure}

In Fig.~\ref{TwB2},  
we present results for the  
singlet-triplet crossover at smaller RKKY coupling,  
for different magnetic fields and finite temperature. Compared to 
the sharp transition in Fig.~\ref{TwB} 
where $J$ is much larger 
than $T_k$, the crossover here is smoother and closer to 
the experimental observation in Ref.~\onlinecite{Heersche}. 
Note that due to thermal broadening at a finite temperature, 
$T(\omega=0)>0$ even in the absence of a magnetic field. 
For antiferromagnetic RKKY coupling (see Fig.~\ref{TwB2}(a)), 
$T(\omega)$ shows the dip-peak-dip structure 
with increasing magnetic field. For ferromagnetic RKKY 
(see Fig.~\ref{TwB2}(b)), the Kondo 
peak at $B=0$, which is due to complete screening of the triplet, 
splits monotonically with increasing field due to Zeeman splitting of 
the triplet levels as shown 
in Ref.~\onlinecite{Heersche}. Both results 
are qualitatively in good agreement with experiment  
\cite{Heersche} (for new measurements on the magnetic field dependence see \cite{Osorio}). 

\emph{Conclusions.} We have studied transport properties 
of a double quantum dot system with RKKY interaction. 
Using the Numerical Renormalization Group we have calculated 
the transmission at finite frequency as a function of temperature and magnetic field. 
A quantum phase transition between the Kondo screened phase and a 
local spin-singlet state  
is observed. Moreover we have shown that both finite temperature 
and a magnetic field can restore the Kondo resonance 
in the presence of an RKKY coupling. 
This crossover back into a Kondo screened state 
is in good agreement with recent measurements \cite{Heersche}. 
For the differential conductance in a finite magnetic field we predict a multiple peak structure 
which is yet to be observed in future experiments. 

\emph{Acknowledgements.}
The authors would like to thank M.~Wegewijs, M. Vojta and H.~van der Zant 
for useful discussions and feedback. 
This work is supported by the Center for Functional Nanostructures 
(CFN) Karlsruhe (C.H.C).

\end{document}